# Ergodicity, ergodicity breaking, and temporal evolution of probability density in non-equilibrium – new perspectives by multi-speckle DLS.

Hans Joachim Schöpe

Institute for Applied Physics, Eberhard Karls University Tübingen, Auf der Morgenstelle 10, 72076 Tübingen, Germany.

email: hans-joachim.schoepe@uni-tuebingen.de

**Abstract**

Testing the ergodicity hypothesis through experimentation in real systems is of crucial importance as it challenges a fundamental assumption underlying a wide range of theoretical models of physical and complex systems. High-calibre experiments can demonstrate whether, when and how a real system explores all of its possible states, and serve as a crucial conduit between abstract theoretical frameworks and tangible real-world applications. This approach is intended to guarantee that our descriptions of complex systems remain both physically meaningful and technologically relevant, thereby facilitating a more nuanced and comprehensive understanding. The development of a novel light scattering experiment has enabled us the direct measurement of the full distribution of fluctuations in the particle concentration in colloidal model systems and thus the direct determination of the probability distribution. This, in turn, facilitates the experimental verification of the ergodic hypothesis and the quantification of ergodicity breaking as a function of metastability.

## Introduction

The concepts of ergodicity and ergodicity breaking[1, 2, 3, 4, 5] are of fundamental importance to a wide range of scientific disciplines, including physics of classical and quantum many-body systems[4, 6, 7, 8, 9], astronomy[10], biology and life science[11, 12, 13, 14, 15, 16], material science[17, 18], cognitive science[19], geoscience[20, 21, 22], engineering[23], information theory and machine learning[24, 25, 26] as well as economics[27]; to name but a few. Ergodicity is of pivotal importance in these various scientific disciplines, as it facilitates comprehension of system development, the attainment of equilibrium, and the prediction of long-term behaviour based on short-term observations. Following Boltzmann the ergodic hypothesis posits that the statistics of the equilibrium state are given by a probability distribution of all states compatible with energy conservation, through which the point representing the state in phase space circulates quasi-periodically - every allowable point in phase space is visited quasi-periodically by the system after a sufficiently long time. In experimental terms, this signifies that the mean value of numerous individual measurements obtained from repeated experiments on copies (ensembles) of a system is equivalent to the mean value of a time average from a single, but very long (ideal: infinitely long), experimental run. A measurement taken over an extended time period will demonstrate the statistical distribution in question. In accordance with the central limit theorem, the mean and sum of independently and identically distributed random variables in any distribution approach the normal distribution as the sample size increases. This is synonymous with ergodicity - if each observable satisfies the central limit theorem, then the system is ergodic. In ergodic samples macroscopic observables can be predicted by uniform averaging over all states, forming the basis for all equilibrium thermodynamic calculations. However, the ergodic hypothesis has not yet been fully verified either mathematically or experimentally[5, 28].

If a complex system is out of equilibrium, ergodicity breaking occurs[29, 30] - strictly speaking, the system is no longer ergodic. This is often correlated to symmetry breaking[31]. Ergodicity breaking is characterised by deviations from the conventional Gaussian probability distribution – the central limit theorem is not fulfilled any more. Additionally, the probability density becomes time-dependent. A fundamental challenge in physics is to predict how systems evolve in time when taken out of equilibrium. When it comes to using or developing the "right" theoretical description in these situations[18, 32, 33, 34, 35, 36, 37, 38], the degree of ergodicity breaking is crucial.

However, this is not easily quantifiable and, to the best of my knowledge, has not yet been determined directly in experiments. I am motivated to do so investigating a colloidal model system using multi-speckle dynamic light scattering.

Colloids represent a significant model system for investigating fundamental questions in non-equilibrium many-particle physics. It is evident that, due to their typical length and time scales, the structure and dynamics of these mesoscopic systems can be studied experimentally using light scattering in detail. In comparison with atomic systems, the results obtained in colloids offer superior temporal and spatial resolution. Dynamic light scattering (DLS), also known as photon correlation spectroscopy, has long been established as a powerful technique for characterizing the dynamics of colloidal dispersions, polymers, emulsions, and complex fluids. The technique relies on measuring temporal fluctuations in the intensity of scattered laser light which are caused by the Brownian motion of the particles under investigation. However, conventional single-detector DLS methods are subject to substantial experimental limitations. When colloids are analysed using classical DLS with a point detector, the resulting signal is an average that allows only indirect inferences about the probability distribution of density fluctuations[39, 40, 41]. The development of multi-speckle correlation spectroscopy[42, 43, 44, 45, 46] , also known as multi-speckle dynamic light scattering or multi-speckle diffusing-wave spectroscopy in the multiple scattering limit[47, 48], represents a transformative breakthrough in addressing these fundamental limitations. It can be regarded as a contemporary iteration of DLS, employing area detectors to analyse the fluctuating speckle pattern. One major achievement of multi-speckle techniques is its capacity to simultaneously resolve spatial heterogeneity of dynamics and temporal evolution of non-stationary systems. In order to address this issue, space-resolved setups were specifically introduced[44, 45, 46, 48]. Another significant advancement goes beyond ensemble-averaged correlation functions; multi-speckle techniques enable quantification of the probability distribution of intensity (density) fluctuations themselves, revealing the degree of dynamical heterogeneity in non-ergodic, non-stationary systems[43, 46]. The deployment of distinct detection schemes is contingent upon the specific application and the nature of the research question being addressed [43, 44, 45, 46, 49].

The construction of a novel light scattering apparatus has enabled direct measurement of the probability distribution of density fluctuations. In order to accomplish this, the complete scattering volume that is illuminated by a laser is divided into several thousand volumes of identical size of 7x7x20 $\mu m^3$. Each of these contains an (sub-) ensemble. The autocorrelation function of the scattered light intensity of the individual ensembles enables an analysis of the probability density of the density fluctuations. In the following, I present measurements on a colloidal hard sphere (HS) system in and out of equilibrium, with which I verify the ergodic hypothesis and quantify the degree of ergodicity breaking as a function of metastability and waiting time.

**Materials and Methods**

For the measurements, I used colloidal particles with HS like interaction[50]. The freezing and melting concentration are scaled to the corresponding values of a monodisperse hard sphere system: $\Phi_{freeze}$ = 0.494 and $\Phi_{melt}$ = 0.545. For my study I used a home build light scattering apparatus consisting of a multi-speckle correlation spectroscopy and conventional dynamic light scattering setup to investigate the particle dynamics and a time-resolved multi-angle Bragg scattering setup to investigate the solidification kinetics. The information from the spatio-temporal fluctuations of the scattering particles is analyzed by calculating the correlation functions of the scattered light intensity field at a scattering vector q. The normalized coherent *time averaged* intensity auto-correlation function (IACF) $g_T^{(2)}(\mathbf{q},\tau,t_0)$ is obtained by recording the time trace of the scattered intensity of a single Fourier component. The duration T of the measurement is several orders of magnitude longer than the correlation time $\tau$ that is of interest. A measurement of several thousands independent spatial Fourier components Sp gives access to the normalized coherent *ensemble averaged* IACF $g_E^{(2)}(\mathbf{q},\tau,t_0)$. In order to obtain an ensemble average, it is a prerequisite that the probed scattering volume contains a sufficiently large number of statistically independent configurations. Specifically, this means that the dynamic correlation length (spatial correlation in the dynamics) in the observed subensemble must be significantly smaller than the distance between the selected subensembles. The ergodic hypothesis assumes that time $\langle\ldots\rangle_T$ and ensemble averages $\langle\ldots\rangle_E$ of the chosen observable are equivalent.

$$g_T^{(2)}(\mathbf{q},\tau) \overset{\text{ergodic hypothesis}}{\equiv} g_E^{(2)}(\mathbf{q},\tau) \tag{1}$$

To obtain meaningful data in non-ergodic systems a combination of time and ensemble averages known as “brute force averaging” $g_{T,E}^{(2)}(\mathbf{q},\tau,t_0)$ promises very good statistics[51]. However, unbiased data can only be obtained with this method if the changes in the dynamics due to aging in the measurement time T are negligibly small. The intermediate scattering function (ISF) $f(\mathbf{q},\tau,t_0)$ is obtained using the Siegert relation. Further details of the experimental setup and the procedures are given in the SI.

## Results

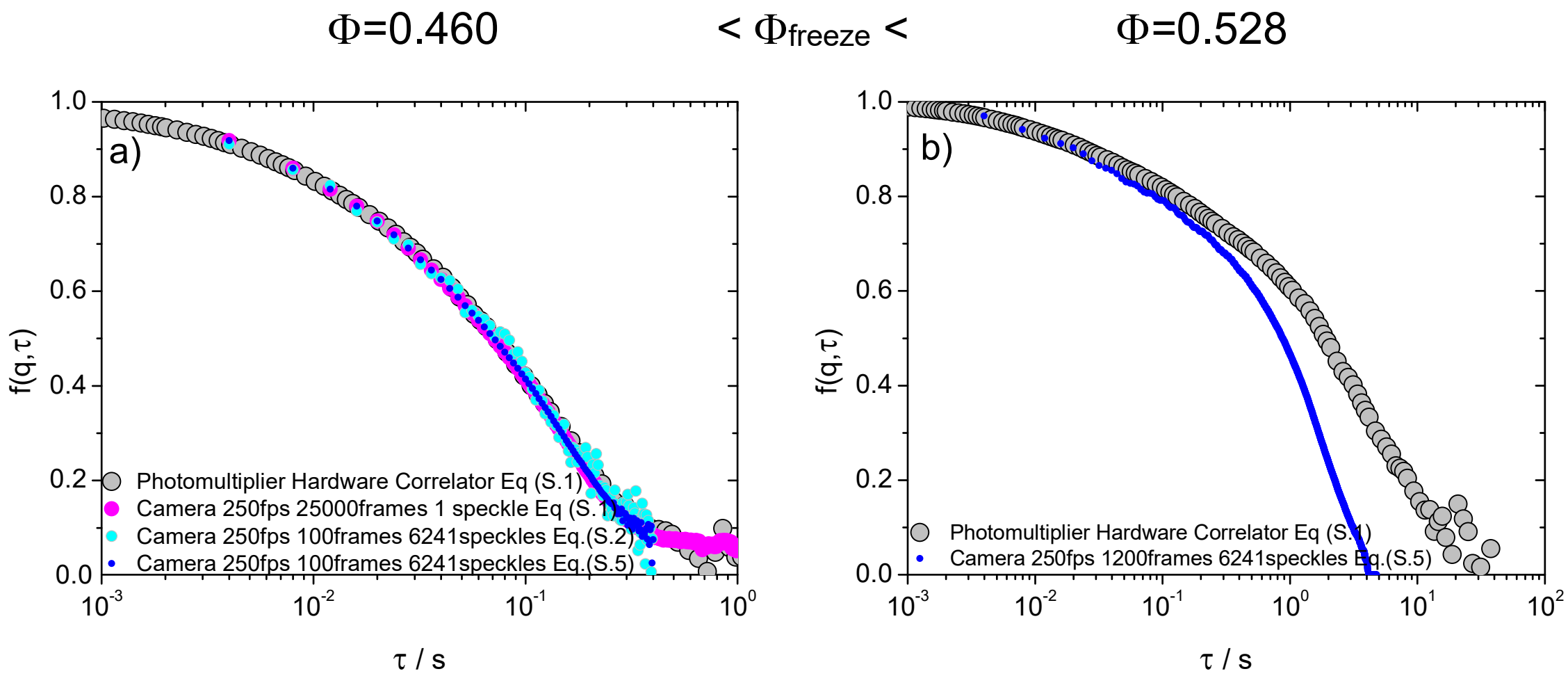

**Fig. 1. Intermediate scattering function in and out of equilibrium.** a) ISF of an equilibrium sample (Φ=0.46) measured by fibre coupled DLS (grey filled circles) and by multi-speckle DLS (pink, cyan and blue filled circles) for various speckle number, measuring times T and averaging (blue, pink: time averaged; cyan: ensemble averaged, blue: brute force averaged – please see SI for more details). Grey circles: T=180s, 10 runs averaged, analysed using Eq (S.1), pink discs: 1 speckle, T=1000s, analysed using Eq (S.1), cyan discs: 6241 speckle, T=$\tau_{max}$=0.4s, analysed using Eq (S.2), blue discs: 6241 speckle, T=$\tau_{max}$=0.4s, analysed using Eq (S.5).
b) ISF of a sample out of equilibrium (Φ=0.528) - grey filled circles: fibre coupled DLS, T=600s, 10 runs averaged, analysed using Eq (S.1)); blue discs: multi-speckle DLS 6241 speckle, T=$\tau_{max}$=5s, analysed using Eq (S.5).

In the event of the colloidal HS suspension being in thermodynamic equilibrium ($\Phi < \Phi_{freeze}$), the ergodic hypothesis should be applicable, with equation (1) being satisfied. In order to verify this, the IACF was determined both in time average $\langle\ldots\rangle_T$ and ensemble averages $\langle\ldots\rangle_E$. The measurements were performed at a spatial frequency corresponding to the position of the principal static structure factor peak q* (averaged particle distance) and in VV-geometry. In figure 1a the resulting ISFs are presented. The time average data (illustrated by grey and pink filled circles) are equivalent to the ensemble averaged data (illustrated by cyan filled circles), thereby providing substantial confirmation of the ergodic hypothesis.

Furthermore, Figure 1a presents a data set (illustrated as blue filled circles) that combines time and ensemble averages $g_{T,E}^{(2)}(\mathbf{q},\tau,t_0)$ at minimum measurement time T=$\tau_{max}$. This evaluation method provides optimal statistics with minimal measurement time, making it ideal for determining the ISF in samples outside of thermodynamic equilibrium. Out of equilibrium (figure 1b) the results of time average and ensemble average do not agree with each other – during the significantly larger measurement period for the time average a temporal evolution of the particle dynamics is observed.

In order to gain a more profound understanding of the dynamics of the metastable melt, it is imperative to directly measure the distribution of relaxation processes. For this purpose, I systematically measured the structure and dynamics of colloidal HS in thermodynamic equilibrium and in the metastable state as a function of volume fraction using the new light scattering apparatus. All measurements were performed at q*. In order to obtain meaningful dynamic data, a minimum of seven measurements were executed at a single volume fraction. As illustrated in Figure 2, the measurement results for a sample in thermodynamic equilibrium with Φ=0.460 (left) and in a metastable state with Φ=0.528 (right) are shown. It is noted that the samples concentration is approximately 10% away from the freezing transition on both sides. The upper graphs a), b) illustrate the ISF, the time auto-correlation function of the longitudinal particle current (CAF), and the negative natural logarithm of the ISF as a measure of a "collective mean squared displacement " as a function of the correlation time. Due to their informative value the collective mean squared displacement and the CAF are shown in a double logarithmic plot. The next row (graphs c), d)) illustrate the corresponding frequency distributions of 6241 intensity autocorrelation functions, colour-coded for clarity. The one-sigma width of the distribution is represented by the yellow colour. The mean value of the 15.85% of IACFs that exhibit the most rapid or least rapid decorrelation is represented by the transition from turquoise to blue. In the third row (graphs e), f)) the relaxation time distribution is shown for a correlation value at which the ISF has dropped to 1/e. The relaxation time distribution is obtained by cutting the frequency distribution of the IACF horizontally at a fixed correlation value. As demonstrated in the graphs e), f) located at the bottom of the presentation, the spatial distribution of density relaxation is finally revealed. The utilisation of colour-coded maps facilitates the representation of the distribution of the normalised IACF, as demonstrated at a specific delay time at which the averaged value has been reduced to 0.5.

In thermodynamic equilibrium, the ISF and CAF can be described with a stretched exponential represented by the blue curve, and the collective mean squared displacement shows a power law $\sim\tau^1$, as is usual for diffusion. The frequency distribution of the IACF manifests as a narrow, symmetrical distribution. In thermodynamic equilibrium ($\Phi$=0.460) the relaxation time distribution is well described by a Gaussian (blue line) demonstrating ergodic behavior. Since a decaying IACF at can be approximated locally using a linear Taylor expansion, their distribution (dynamic susceptibility) is also given by a Gaussian function (vertical cut of the frequency distribution of the IACFs). The spatial distribution of the density relaxations exhibits only minor fluctuations.

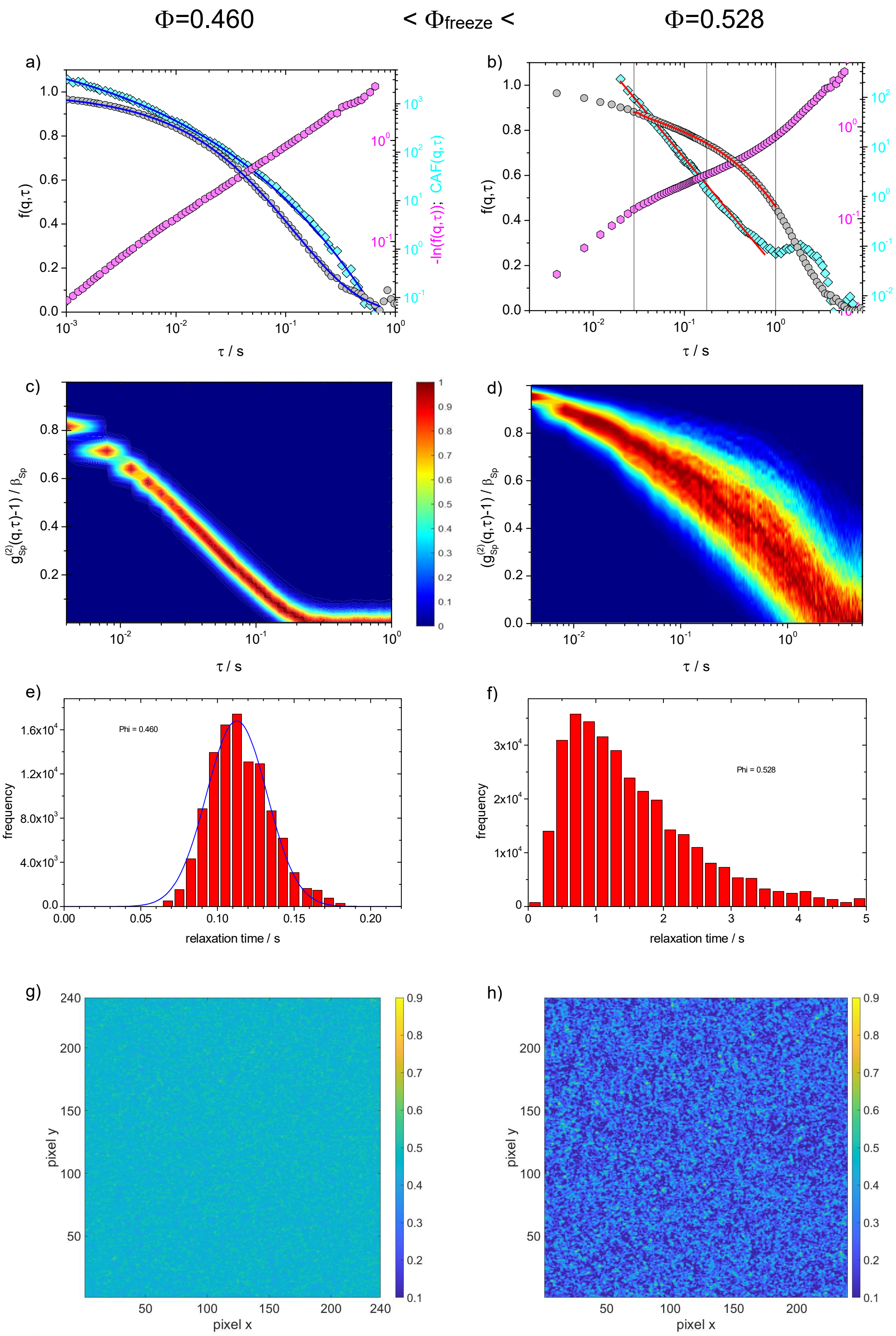

**Fig. 2. Spatio-temporal particle dynamics in and out of equilibrium**. Top a), b): ISF f(q,τ), CAF C(q,τ) and "collective mean squared displacement" –ln(f(q,τ)) for a samples in equilibrium (Φ=0.460, left) and in the metastable state (Φ=0.528, right). The blue lines are stretched exponentials; the red lines power laws, fitted to the ISF and CAF. Middle top c), d): Normalized frequency distribution of the corresponding normalized time-averaged ISFs. The color-coded diagrams show the frequency distribution from 6241 individual autocorrelation functions. Bottom top e), f): Frequency distribution of the relaxation time at f(q,$\tau$)=1/e. The blue curve is a Gaussian curve that has been fitted to the data ($\Phi$=0.460). Bottom g), h): Activity maps showing the spacial distribution of normalized time-averaged ISFs at a delay time where $\left\langle (g^{(2)}(\mathbf{q},\tau,t_0)-1)/\beta) \right\rangle = 0.5$

Out of equilibrium, the collective mean squared displacement in double logarithmic plotting demonstrates an inflection point at approximately 0.2s and, in the limit, at short ($\tau$<0.03s) and long ($\tau$>1s) correlation times a linear course (power law). This behaviour is indicative of the cage effect (anomalous diffusion) [40]. Accordingly, the ISF and CAF cannot be described with a stretched exponential decay in the time window of the cage effect, but can be well described with a power law represented by the red curves [35]. This is readily apparent in the double logarithmic plot of the CAF. The frequency distribution of the ISF demonstrates a broad asymmetric distribution. It is noteworthy that the rapidly decorrelating IACFs (illustrated by the transition from turquoise to blue at short times) demonstrate an exponential decay to a good approximation, while the slowly decorrelating IACFs (illustrated by the transition from turquoise to blue at long times) exhibit a plateau in the middle time range (~0.2 s) prior to a rapid decay at the end of the time window. It is evident that these data represent the cornerstones of the heterogeneous spatio-temporal dynamics, i.e. dynamic heterogeneities, which are caused by fluid and solid-like regions [46, 52, 53]. The spatial distribution of density relaxation, as exhibited in the dynamic map, manifests the presence of solid-like regions (yellow and orange) encircled by fluid. The characterisation of such dynamic heterogeneities frequently involves the use of the dynamic four-point susceptibility function $\chi_4$ [43, 54, 55, 56, 57]. The standard deviation of the autocorrelation function distribution (yellow to yellow in fig.2) directly reflects the susceptibility: $\sigma(f(q,\tau)) = \chi_4$. In the case of the metastable fluid with $\Phi$=0.528, $\chi_4$ initially increases continuously with increasing correlation time and reaches its maximum at a correlation time of approximately 0.8 s. This is the point at which structure and dynamics are most intimately intertwined and coincides, as a good approximation, with the end of the caging (anomalous diffusion) time window. Subsequently, the susceptibility decreases steadily until the end of measurement, which reflects qualitatively the well-known pattern[56].

The relaxation time distribution in non-equilibrium exhibits significant deviations from a Gaussian. The distinction between ergodic (Φ=0.460) and non-ergodic (Φ=0.528) is clearly evident. In order to specify the transition from Gaussian to non-Gaussian fluctuations and thus the degree of ergodicity breaking, the central moments $m_i(\tau_R) = \left\langle \left( \tau_{R,Sp} - \left\langle \tau_{R,Sp} \right\rangle \right)^i \right\rangle$ of the measured relaxation time distributions were determined. $\tau_{R,Sp}$ denotes the relaxation time of a single speckle.

The progression of the first four moments as a function of volume fraction is illustrated in Figure 3. The open symbols indicate the values for the long-term component when a bimodal distribution was identified. The lines are guide lines ("spline fits") to highlight the evolution of the data. The course of the moments demonstrates the transition from a narrow symmetric distribution in thermodynamic equilibrium to a broad asymmetric, partly bimodal distribution in non-equilibrium. Of particular note is the "jump" of the higher moments at the freezing point, marked by the vertical line at Φ=0.494. The transition from Gaussian fluctuations in thermodynamic equilibrium to non-Gaussian fluctuations in the metastable fluid is clearly identifiable. Furthermore, the proportion of non-Gaussian fluctuations increases continuously outside thermodynamic equilibrium with increasing volume fraction (increasing metastability). The observed transition in particle dynamics "at" the freezing point thus reflects the transition from an ergodic system to a non-ergodic one. With this finding, these results corroborate the interpretation of previous light scattering studies on HS colloids [41].

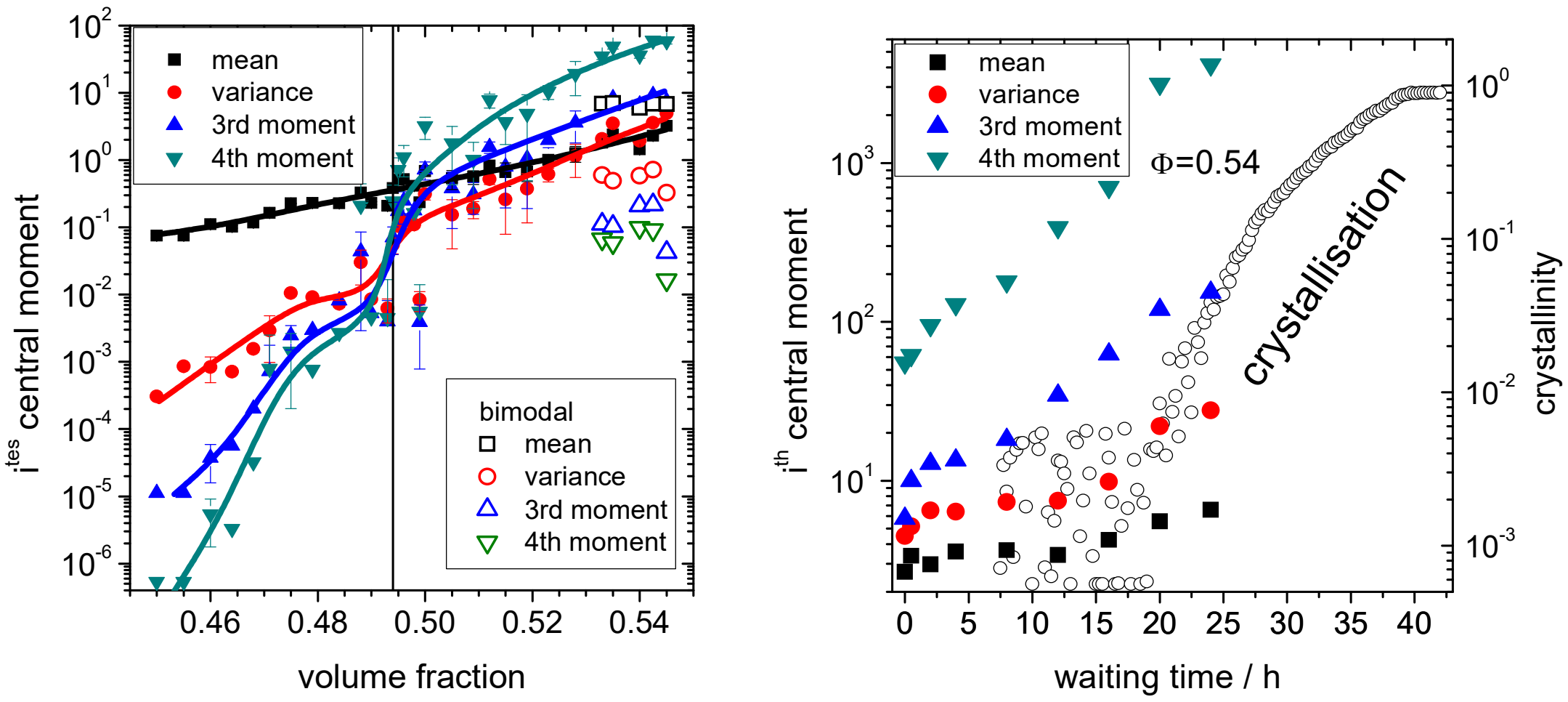

**Fig. 3. Moments of the probability distribution as function of particle concentration and as function of time out of equilibrium.** Left: The first four central

moments of the relaxation time distribution as function of volume fraction. Open symbols describe the long-term component of bimodal distributions, which could be observed for volume fractions larger than 0.528. The lines are splines fits to guide the eye. Right: Waiting time dependence of the first four central moments of the relaxation time distributions in the induction phase before crystallization sets in. The open symbols represent the crystallinity (relative amount of crystals) – please see SI for more details.

The configuration of the light scattering apparatus enables the simultaneous measurement of the temporal evolution of dynamics and statics. By doing so I demonstrate that the temporal evolution of the relaxation time distribution of a crystallising sample in the induction period before crystallisation sets in is not, in fact, stationary when observed in laboratory time[49]. As illustrated in Figure 3, the temporal evolution of the first four moments of the relaxation time distribution and the time trace of the crystallinity (please see SI for more information) are depicted for a sample close to the melting volume fraction. In the induction phase, the proportion of non-Gaussian fluctuations ($3^{rd}$ and $4^{th}$ moments) increases sharply until the onset of crystallisation at approximately 22 hours - the dynamics of the metastable fluid are non-stationary. The proportion of non-Gaussian fluctuations increases continuously over time. Theoretical descriptions of phase transition phenomena that use a Boltzmann statistics should therefore be viewed critically.

**Conclusions**

I interpret my findings as follows. The data presented herein demonstrate a normal distribution of the probability distribution of density fluctuations in the thermodynamic equilibrium fluid. In equilibrium fluid states of matter, the dynamics continually restore stationarity and ergodicity already at the microscopic scale. This is akin to it being Markovian, and a balance of fluxes holds locally, often denoted as detailed balance.

In contrast, for non-equilibrium fluids, this no longer holds. The central limit theorem is not fulfilled and probability densities are time-dependent. Ergodicity breaking occurs, which is intimately related to a break in the symmetry of time and space initially at a local scale (caging). The broken detailed balance leads to heterogeneous spatio-temporal dynamics which can be impressively visualized with the help of multispeckle DLS. In the context of dynamic heterogeneities, it is proposed that the slow non-equilibrium fluctuations in the relaxation time distributions are indicative of the collective particle dynamics in the form of solid-like phonons, which are a consequence of the cage effect. The phonons stabilize the solid like regions mechanically. With increasing metastability the amount of the solid like regions increases and they become stiffer. With increasing waiting time the nonlinear dynamical nature of the system enables a disbalance on the microscopic scale to propagate up to larger, mesoscopic scales of the fluid. The continual action of such irreversible processes ultimately drive the system towards the new equilibrium state. The longer the system remains metastable, the larger the "bandwidth" of collective dynamics that may be able to evolve. This phenomenon is clearly evident in the time-dependent measurements of the probability density.

In summary, it is evident that spatially resolved multi-speckle correlation spectroscopy enables the direct measurement of the probability density of density fluctuations in a colloidal fluid. This facilitated the verification of the ergodic hypothesis and the central limit theorem in the experiment. The transition from equilibrium fluid to metastable fluid is characterised by an abrupt change in the shape of the probability density. As metastability increases, the proportion of non-Gaussian fluctuations increases continuously. In the non-equilibrium state, the probability densities are time-dependent, and the bandwidth of non-Gaussian fluctuations increases with time. The utilisation of spatially resolved multi-speckle correlation spectroscopy facilitates the characterization of dynamic heterogeneities within reciprocal and direct space: regions that bear the dynamic signature of a solid are embedded in a "sea" of liquid. The regions in which the particles execute localised collective movements (comparable to phonons in solids) increase with increasing metastability and with waiting time; furthermore, the cages become stiffer.

**Acknowledgments**

I gratefully acknowledge financial support by the Deutsche Forschungsgemeinschaft (SCHO 1054/7-1). I thank Bill van Megen, Tanja Schilling, Miriam Klopotek and Martin Oettel for fruitful discussions.

**Author Contributions**

HJS designed the project, designed and realized the experimental set up, characterized and prepared samples, performed all measurements, wrote Matlab codes; analyzed experimental data, has written the manuscript.

## Data availability

All data generated and analysed during the current study are stored by the corresponding author and are available upon request.

## Supplementary Information for

# Ergodicity, ergodicity breaking, and temporal evolution of probability density in non-equilibrium – new perspectives by multi-speckle dynamic light scattering.

Hans Joachim Schöpe

email: hans-joachim.schoepe@uni-tuebingen.de

### Experimental procedures

For the measurements, I used colloidal particles with HS-like interaction. The particles consist of a copolymer core of methylmethacrylate and trifluoroethylacrylate, with a stabilizing coating of poly-12-hydroxystearic acid, about 10 nm thick, bonded to the surface[1, 2]. To reduce the suspension's turbidity, the particles and the suspending solvent, cis-decalin, have refractive indices that are closely matched: at a sample concentration of $\Phi$=0.5 the samples display a transmission of ~97% at T=16°C, which corresponds to a turbidity of 2.61 $m^{-1}$. Figure 1 shows a photograph of a sample with $\Phi$=0.5 at 23°C – out of perfect optical match.

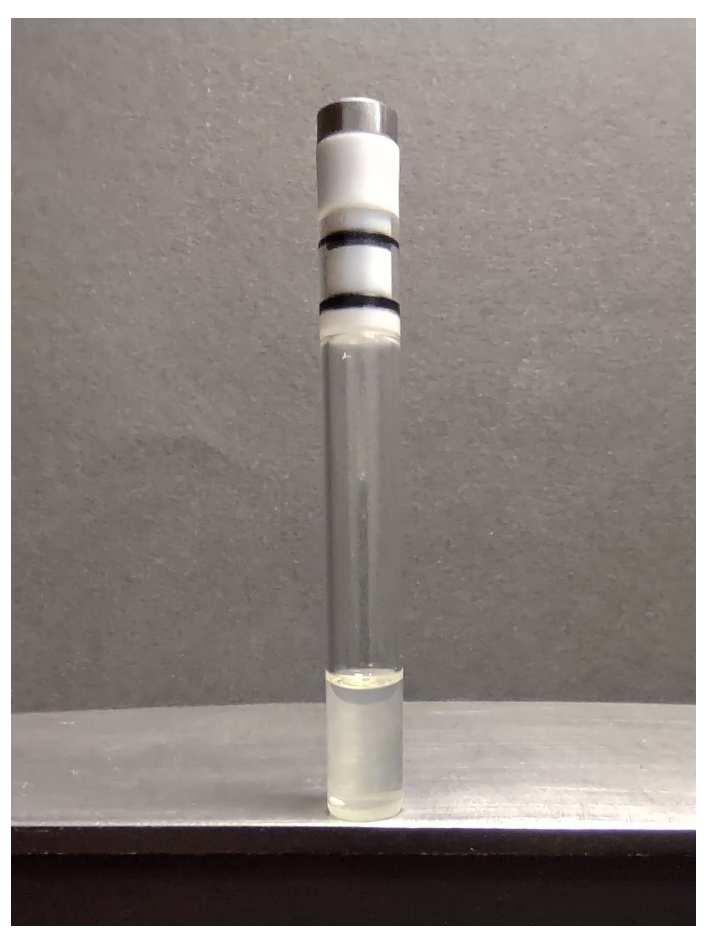

**Fig. S1 Photograph of a sample.**

Hard sphere like colloidal suspension at concentration of $\Phi$=0.5 at room temperature in a cylindrical cuvette.

The particle size distribution was determined by electron microscopy resulting in a nearly symmetric distribution with a mean radius of $R_{EM}$ = 180nm and a size polydispersity 7%. The hydrodynamic radius was determined to be $R_{DLS}$=187nm, which leads to a Brownian time $t_B$ = $R^2/6D_0 \approx 0.017$ s, where $D_0$ is the free particle diffusion coefficient. The freezing and melting concentration analyzing the equilibrium phase behavior [3] and are scaled for simplicity to the corresponding values of a monodisperse hard sphere system: $\Phi_{freeze}$ = 0.494 and $\Phi_{melt}$ = 0.545.

All samples are tumbled for at least 12h before starting a light scattering (LS) measurement. After cession of shear the sample was directly mounted into the LS-setup and I waited 30 min for calibration before I started a measurement. For my study I used a home build LS - apparatus consisting of a multispeckle correlation spectroscopy (MSCS) and conventional dynamic light scattering (DLS) setup to investigate the particle dynamics and a time-resolved multi-angle Bragg scattering setup (TRSLS) to investigate the solidification kinetics. This LS apparatus is an entirely new design based on my experience with the apparatus we used to investigate structural and dynamic heterogeneities at the glass transition[4]. The set up therefore features, among other things, two independently movable goniometer-arms, highly reproducible sample positioning (in the low µm range), exceptionally good long-term stability, optimized optics for the measurement methods used, and in multispeckle correlation a remarkable spatial resolution. For MSCS measurements, I used a camera framerate of 250 fps and recorded up to 50000 frames. The movies where analysed using self written Matlab programs. Up to 6241 individual IACFs were measured. The light from the individual IACFs originates from separated sample volumes of 7x7x20 µm$^3$ each in size.

**Dynamic light scattering**

The information from the spatio-temporal fluctuations of the scattering particles is analysed by calculating the correlation functions of the scattered light intensity field of a given Fourier component q. It should be noted that, in a real light-scattering experiment, there is an unavoidable q-smearing, which depends on the optics used. Within the confines of the used configuration, this is equivalent to approximately dq/q ≈ 0.8%. The normalized coherent *time averaged* intensity auto-correlation function (IACF) is given by

$$g_T^{(2)}(\mathbf{q},\tau,t_0) = \frac{\lim\limits_{T\to\infty}\frac{1}{T}\int\limits_{t=t_0}^{t_0+T} I(\mathbf{q},t)\, I(\mathbf{q},t+\tau)\, dt}{\langle I\rangle_T^2} \equiv \frac{\langle I(\mathbf{q},t_0)\, I(\mathbf{q},t_0+\tau)\rangle_T}{\langle I\rangle_T^2} \qquad \text{(S.1)}$$

, where $\tau$ is the correlation time and T the duration of the measurement. T is several orders of magnitude longer than the correlation time that is of interest. A measurement of independent spatial Fourier components Sp of the fluctuations offers the possibility to calculate the normalized coherent *ensemble averaged* IACF:

$$g_E^{(2)}(\mathbf{q},\tau,t_0)=\frac{\frac{1}{N}\sum_{Sp=1}^{N\to\infty} I_{Sp}(\mathbf{q},t_0)\, I_{Sp}(\mathbf{q},t_0+\tau)}{\langle I(\mathbf{q},t_0)\rangle_E \langle I(\mathbf{q},t_0+\tau)\rangle_E}=\frac{\langle I(\mathbf{q},t_0)\, I(\mathbf{q},t_0+\tau)\rangle_E}{\langle I(\mathbf{q},t_0)\rangle_E \langle I(\mathbf{q},t_0+\tau)\rangle_E} \qquad \text{(S.2)}$$

Experimentally estimates of Eq. (S.2) can be obtained by averaging over a very large number (several thousand) of independent fluctuating laser speckles. To obtain an ensemble average, it is a prerequisite that the experiment probes a very large number of statistically independent configurations — that is, the scattering volume must divided in a sufficiently large number of subensembles. In the performed experiment the density fluctuations of ~$10^8$ particles are analyzed located in 6241 subensembles. Please note that only two measurements at times $t_0$ and $t_0+\tau$ are multiplicated. That is why $g_E$ is also called "two time correlation function"[5]. The ergodic hypothesis assumes that time $\langle\ldots\rangle_T$ and ensemble averages $\langle\ldots\rangle_E$ of the chosen observable are equivalent.

$$g_T^{(2)}(\mathbf{q},\tau) \overset{\text{ergodic hypothesis}}{\equiv} g_E^{(2)}(\mathbf{q},\tau) \qquad \text{(S.3)}$$

Since ergodicity und thus time isotropy is given in thermodynamic (TD) equilibrium the auto-correlation function is only a function of the lag time $\tau$ - the start time $t_0$ is not important. For a complex random Gaussian field the intensity correlation function and the field correlation function are connected via the Siegert relation

$$g^{(2)}(\mathbf{q},\tau)=1+\beta\left|g^{(1)}(\mathbf{q},\tau)\right|^2 \qquad \text{(S.4)}$$

, where $\beta=g^{(2)}(q, 0)-1$ is the intercept of the IACF. To obtain meaningful data in non-ergodic systems somehow a combination of Eq. (S.1) and (S.2) known as "brute force averaging" promises very good statistics, since it represents an average over independent Fourier components and time[6]. However, unbiased data can only be obtained with this method if the changes in the dynamics due to aging in the measurement time T are negligibly small.

$$g_{T,E}^{(2)}(\mathbf{q},\tau,t_0)=\frac{\frac{1}{N}\sum_{Sp=1}^{N}\langle I_{Sp}(\mathbf{q},t_0)\, I_{Sp}(\mathbf{q},t_0+\tau)\rangle_T}{\left(\frac{1}{N}\sum_{Sp=1}^{N}\langle I_{Sp}(\mathbf{q},t_0)\rangle_T\right)^2}=\frac{\langle I(\mathbf{q},t_0)\, I(\mathbf{q},t_0+\tau)\rangle_{T,E}}{\left(\langle I(\mathbf{q},t_0)\rangle_{T,E}\right)^2} \qquad \text{(S.5)}$$

The normalized field auto-correlation (FACF) function can be identified with the normalized intermediate scattering function (ISF) $f(\mathbf{q},\tau)$ and is defined as

$$g^{(1)}(\mathbf{q},\tau) = f(\mathbf{q},\tau) = \frac{\langle \rho(\mathbf{q},0)\rho^*(\mathbf{q},\tau)\rangle_E}{\left\langle |\rho(\mathbf{q})|^2 \right\rangle_E} \quad \text{(S.6)}$$

, containing all information about the particle dynamics and thus the corresponding propabilitiy densities. Here

$$\rho(\mathbf{q},t) = \sum_{j=1}^{N} \exp(-i\mathbf{q}(\mathbf{r}_j(\mathrm{t})) \quad \text{(S.7)}$$

is the $q^{th}$ spatial Fourier component of the number density $\rho$ , $\mathbf{r}_j(t)$ the position of particle j at time t and * denotes the complex conjugate.

The time auto-correlation function of the longitudinal particle current (CAF) is given by

$$C(\mathbf{q},\tau) = \mathbf{q}^2 \langle j_L(\mathbf{q},0)\, j_L^*(\mathbf{q},\tau)\rangle / S(\mathbf{q}) = -\left\langle \frac{d}{d\tau} n_P(\mathbf{q},0) \frac{d}{d\tau} n_P^*(\mathbf{q},\tau) \right\rangle \Big/ S(\mathbf{q}) = -\frac{d^2}{d\tau^2} f(\mathbf{q},\tau) \quad \text{(S.8)}$$

It contains additional information about the collective particle dynamic, where the particle current is given by

$$j_L(\mathbf{q},t) = \frac{1}{\sqrt{N}} \sum_{j=1}^{N} \hat{\mathbf{q}}\mathbf{v}_j(t) \exp(i\mathbf{q}(\mathbf{r}_j(t)) \quad \text{(S.9)}$$

and $v_j(t)$ is the velocity of particle j at time t.

**Experimental setup and data analysis in MSCS**

The new multipurpose light scattering setup combines a conventional static light scattering (SLS) experiment with time-resolved multi-angle scattering setup (TRSLS) to investigate the temporal evolution of the structure and a conventional dynamic light scattering (DLS) experiment with a multispeckle correlation spectroscopy (MSCS) setup to investigate the particle dynamics. Figure 2 shows a schematic and photographs of the new light scattering set up.

In the SLS configuration, the sample is exposed to laser light with a wavelength of 632nm (Research Electro-Optics Inc.). The laser light is guided onto the goniometer arm using a single-mode fibre. The light passes through a beam expander, a rectangular aperture, and finally a cylindrical lens. This results in a large beam size in the sample, which measures 4mm in width and 10mm in height. The sample itself is mounted in the center of an index match bath which is temperature controlled. For the temperature control we use a thermostat (Thermo Scientific) and the index-match-bath is a customization from Hellma GmbH. The scattered light passes through a rectangular aperture of $4x10mm^2$, then it is coupled into a multimode fibre and detected using a photomultiplier (Hamamatsu) for conventional SLS and simultaneously focused onto a CCD camera (Hamamatsu) for TRSLS using two cylindrical lenses. The CCD-camera detects a scattering angle range of ~12°.

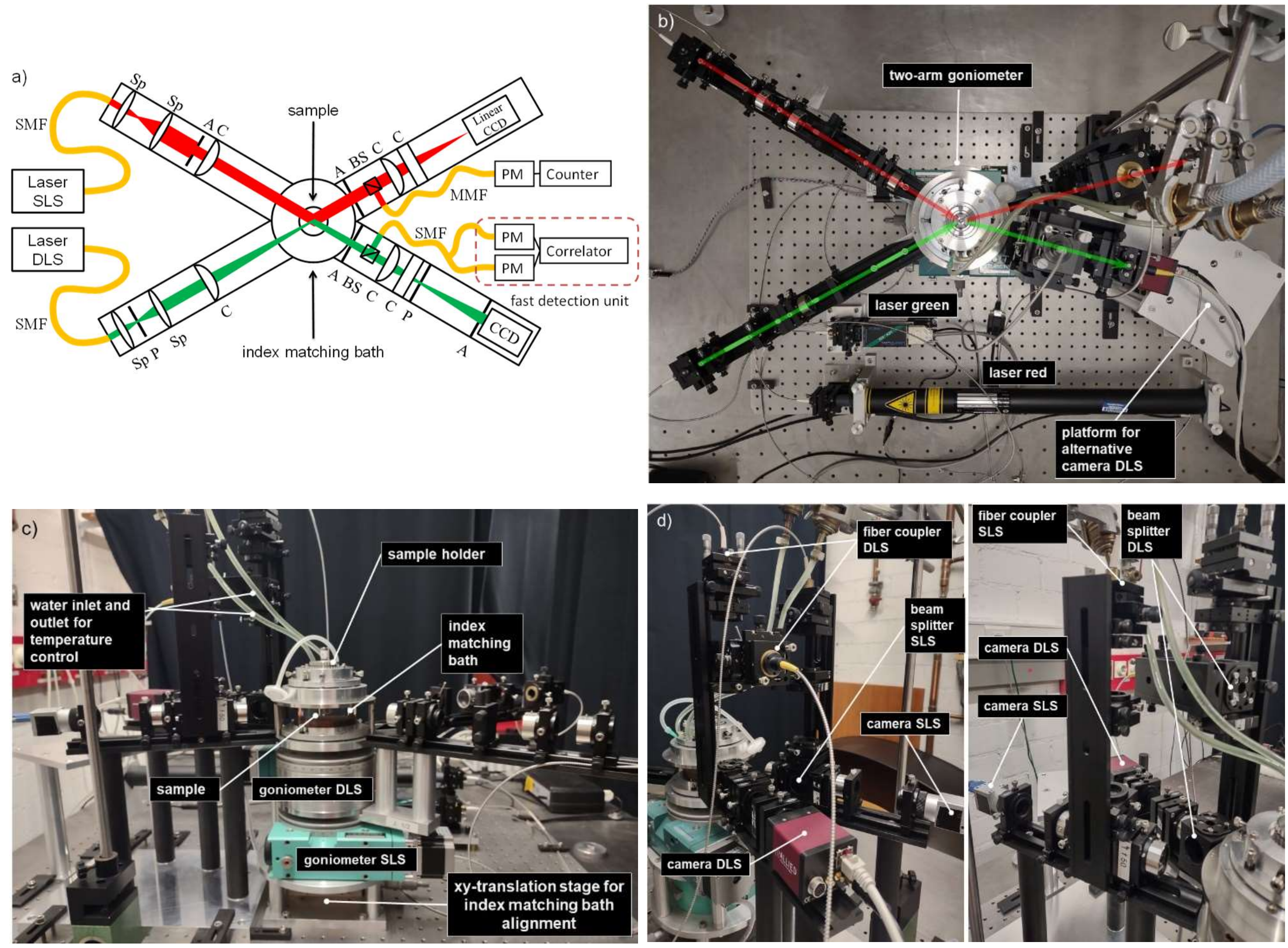

**Fig. S2 Multipurpose light scattering apparatus.**

a) scematic: The sample is located in a temperature controlled index-matched bath. For SLS (red beam path), a large sample volume (4x4x10mm$^3$) is used. The scattered light can be detected simultaneously with a photomultiplier or a CCD camera. In DLS, the sample is illuminated with a light sheet (cross-section: 20µm x 4mm). The scattered light can be analyzed using two photomultipliers and a hardware correlator or using a CCD camera and software correlator. For more details, please see text. Photograph in b) top and in c) side view. Photograph d) shows the detection optics from two different perspectives.

For the DLS and MSCS, the sample is illuminated using laser light of a wavelength of 532nm (Oxxius S.A.). Once more, the laser light is directed onto the goniometer arm through the utilisation of a single-mode fibre. The light passes through a beam expander and is subsequently focused into the sample by means of a cylindrical lens. The beam width of the sample is 20µm, with a beam height of 4mm. The scattered light passes through a rectangular aperture of $2x4mm^2$, is coupled into a single-mode fibre and detected using two photomultiplier (Hamamatsu) for conventional DLS. The autocorrelation function is calculated using a hard ware correlator (ALV-GmbH). At the same time, the light is directed towards a fast CCD camera (AVT Prosilica GE 680). Alternatively, other cameras (e.g., Hamamatsu Orca) can be used. Two different laser wavelengths and interference filters in front of the detectors enable static and dynamic measurements to be performed simultaneously.

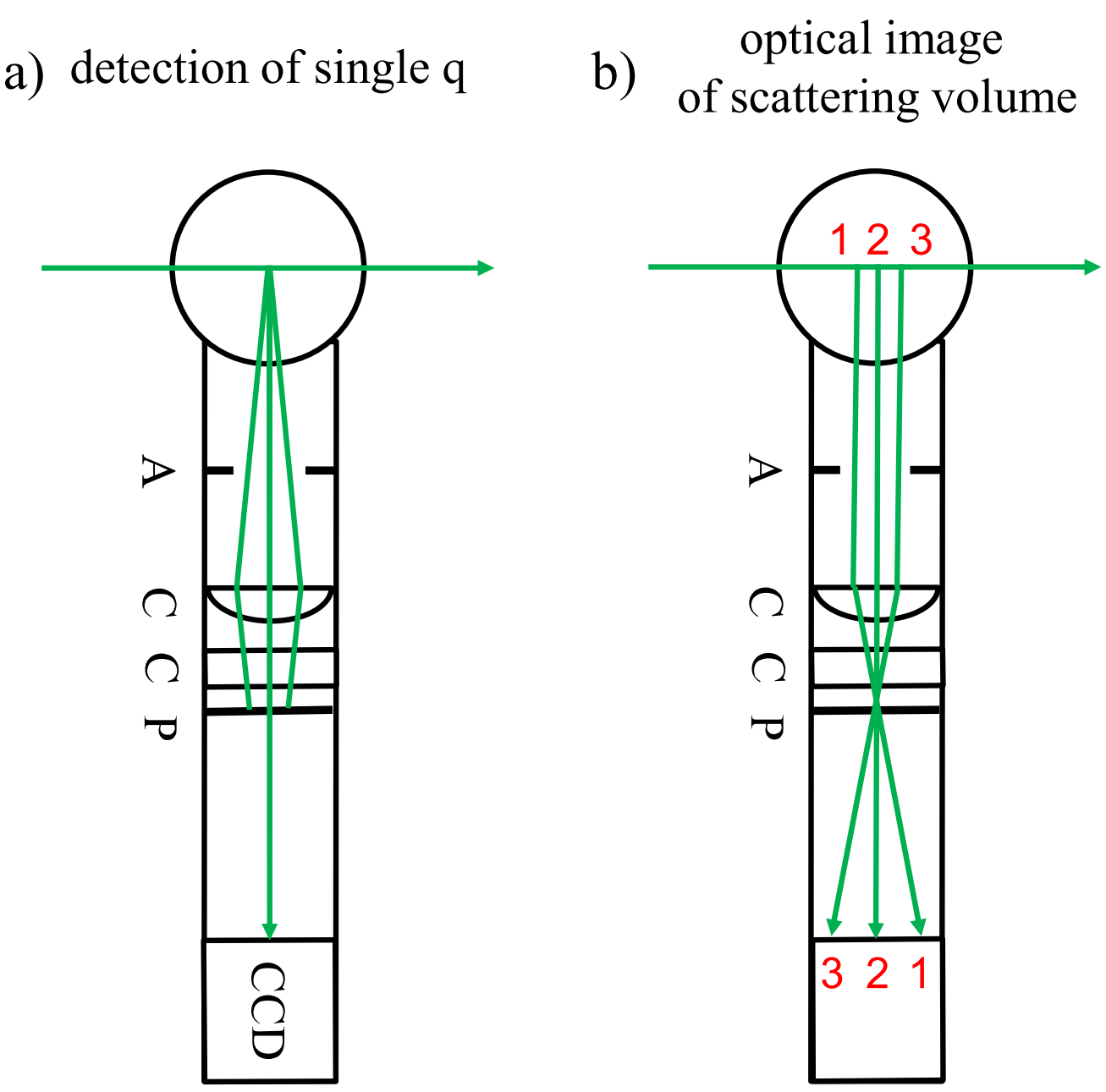

**Fig. S3 Schematic of the detection scheme in MSCS.**

a) Due to the presence of a pinhole (P) in the focal point of the two cylindrical lenses (C), only the scattered light parallel to the optical axis is detected. This guarantees the q-selection. As illustrated in image b), the position on the CCD can be connected to the origin of the scattered light. This allows the analysis of different ensembles.

Figure 3 presents the detection scheme of the MSCS setup. The optical system guiding the light onto the CCD camera ensures that the scattering volume is directly imaged onto the CCD camera. By focusing the scattered light with two perpendicular cylindrical lenses through a pinhole, it is guaranteed that only light scattered under the identical scattering vector q is detected. With the direct imaging, every speckle on the CCD camera is related to its own scattering volume. Therefore, the dynamics at different sample positions and thus in several thousand different ensembles can be investigated simultaneously.
The striking feature of the setup is that more than 6000 intensity autocorrelation functions (IACF) can be measured simultaneously under an identical scattering vector and assigned to clearly separated sample volumes measuring 7x7x20μm$^3$ in size. In “conventional” multispeckle correlation spectroscopy (DLS and XPCS) [7, 8], each coherence area (speckle) is generated from the entire illuminated sample volume (DLS), or the entire coherent volume (XPCS). Different speckles therefore contain redundant information, which makes it impossible to directly measure the probability distribution of dynamic quantities. Our setup is significantly different in that it allows for the direct analysis of dynamic susceptibility as independent copies (ensembles) of the system of interest are probed. To ensure this, the camera signal is first characterized as follows.

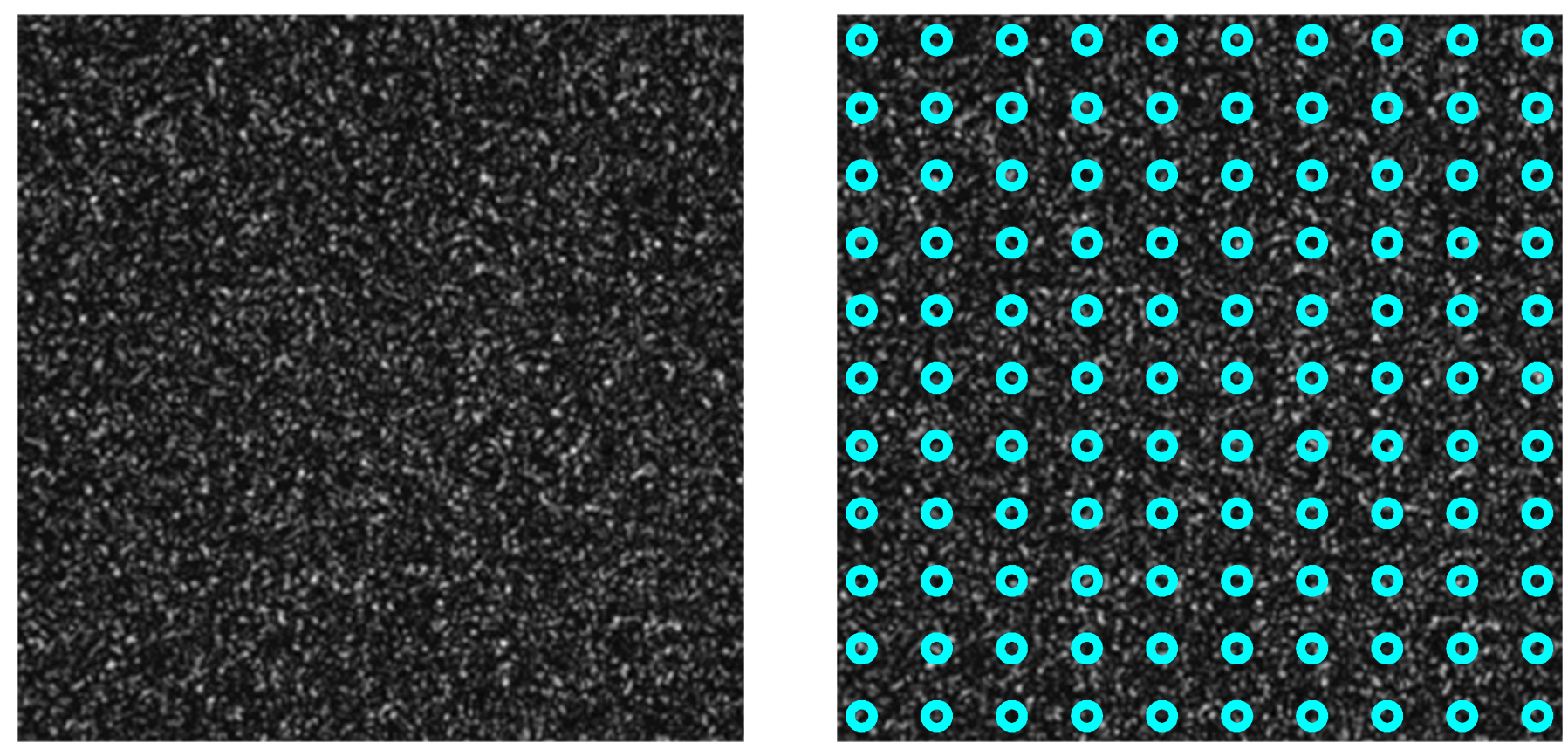

**Figure S4**
Left: Example of a speckle pattern detected by the camera; right: Speckles are read from a square pattern.

The size of the speckle on the camera is initially determined in two ways. First, the image is converted to a black-and-white image and the average speckle size is determined using image analysis. Secondly, the image is subjected to Fourier transformation, thus enabling the calculation of its mean frequency. In order to ascertain the extent to which neighbouring camera pixels contain redundant information, a correlation function of an image series is calculated in time and space. Images that are shifted in time (correlation time $\tau$), are shifted horizontally or vertically, pixel by pixel (correlation length L). In the absence of redundant signal, the correlation function approaches zero. When the equipment is appropriately calibrated, the correlation length is equivalent to the speckle size. In order to ascertain the precise determination of independent IACFs, the readout of speckles exhibiting a minimum distance from each other is undertaken. With a minimum distance of 3 speckles, 6241 IACFs are determined. Figure 4 shows, on the left, an example of a speckle pattern detected by the camera; on the right, the readout scheme is shown schematically. The circles represent the pixels from which the intensity is read out. Speckles are read from a square pattern. For example, if 6241 speckles are read, they are each 3 pixels apart; if 49 speckles are read, they are 30 pixels apart. If only one speckle is read, it is located in the center of the camera. In a final measurement the fluctuating speckle pattern is recorded. The avi file is converted into a MatFile, enabling individual pixels to be addressed at specific times using Matlab. The intensity of the selected speckles is determined as a function of time. The IACF is calculated by a software correlator in Matlab.

**Dynamic susceptibility as function of speckle number analysed**

One might argue that the spread in the measured density fluctuations might depend on the number of speckles analysed. Figure 5 shows the measured distribution of $(g_{Sp}{}^{(2)}(\mathbf{q},\tau,t_0)-1)/\beta_{Sp})$ at a delay time $\tau$ where $\langle (g^{(2)}(\mathbf{q},\tau,t_0)-1)/\beta) \rangle = 0.5$ and the width of a Gaussian fitted to the data as function of speckle number. The sample concentration is $\Phi$=0.46. In the experiment, there is no such relationship.

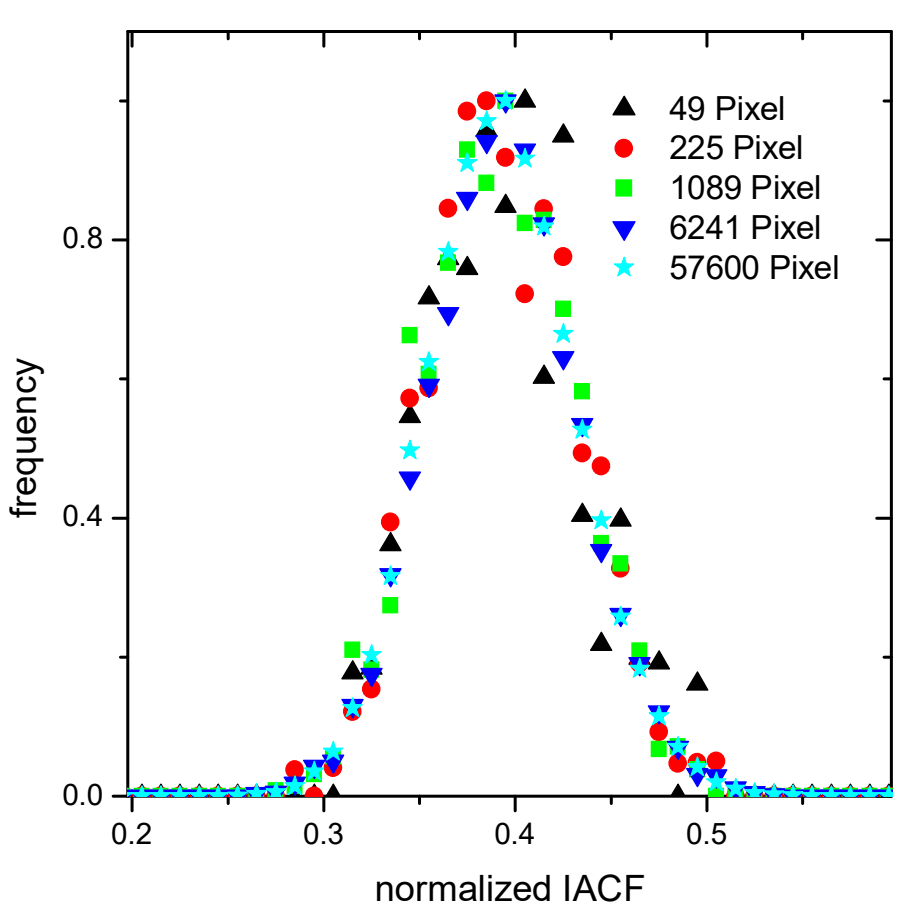

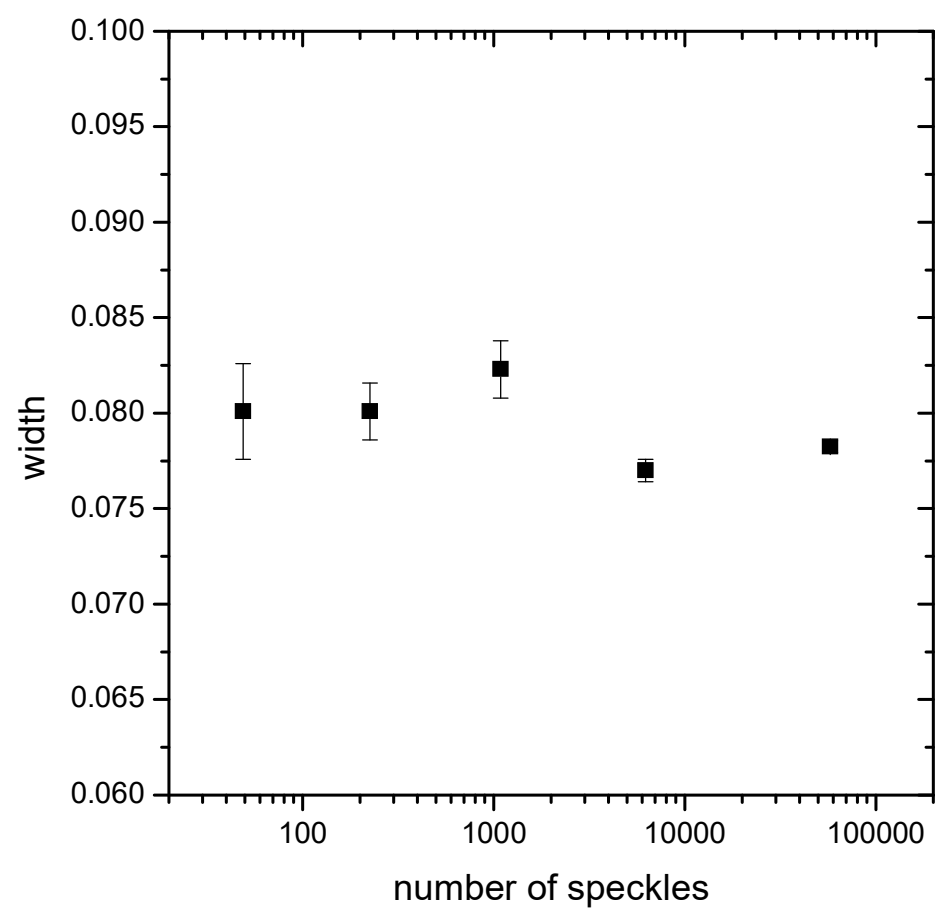

**Figure S5**

Left: Normalized frequency distribution of $(g_{Sp}{}^{(2)}(\mathbf{q},\tau,t_0)-1)/\beta_{Sp})$ for various speckle numbers as indicated at a delay time where the mean dropped down to 0.5. Right: Width of a Gaussian function fitted to the data as a function of the number of speckles.

### Data analysis in TRSLS – determination of crystallinity

The solidification process is characterized by the temporal evolution of the solid static structure factor $\alpha(t_{wait})S_{solid}(q, t_{wait})$, which is obtained by subtracting the fluid background $S_{fluid}(q,t_{wait}) = \beta(t_{wait})S(q,t_{start})$. This procedure was first introduced to colloidal science by Harland and van Megen [9]. $\beta(t_{wait})$ is a scaling factor accounting for the scattering from the decreasing amount of fluid and $\alpha(t_{wait})$ accounts for the increasing amount of solid material. $t_{wait}$ denotes the time passed since cessation of shear melting. A more detailed description of the following described analysis can be found in [10, 11]. For a quantitative analysis of the solidification process, we performed a Pseudo-Voigt fit of the main peak of the solid structure factor (precursor or $(111)_{FCC}$). From the fit the following time dependent parameters were obtained: $A_{hkl}(t_{wait})$ - the area of the peak, $\Delta q_{hkl}(t_{wait})$ – the full width at half maximum (FWHM) and $q_{hkl}(t_{wait})$ – the peak position of the individual Bragg peaks with the Miller indices hkl.
These fit parameters give access to the temporal evolution of several parameters characterizing the solidification process including the crystallinity $X_{hkl}(t_{wait})$ (relative amount of solid material). It is given by the normalized peak area: $X_{hkl}(t_{wait}) = c\, A_{hkl}(t_{wait})$, where c is a normalization factor based on the equilibrium phase diagram.